\documentclass[12pt,a4paper]{article}
\usepackage{amsfonts}
\usepackage{amsmath}
\usepackage{amsbsy}
\usepackage{amssymb}
\usepackage{amstext}
\usepackage{amsopn}

\input{tcilatex}

\begin{document}

\title{A Group-Theoretical Method for Natanzon Potentials in
Position-Dependent Mass Background}
\author{S.-A. Yahiaoui, M.\ Bentaiba\thanks{%
Corresponding author: bentaiba@hotmail.com} \\
LPTHIRM, D\'{e}partement de Physique, Facult\'{e} des Sciences,\\
Universit\'{e} Saad DAHLAB de Blida, Algeria}
\date{}
\maketitle

\begin{abstract}
A new manner for deriving the exact potentials is presented. By making use
of conformal mappings, the general expression of the effective potentials
deduced under $\mathfrak{su}\left( 1,1\right) $\ algebra can be brought back
to the general Natanzon hypergeometric potentials.


\end{abstract}

\section{Introduction\protect\bigskip }

\bigskip Exact solutions for some quantum mechanical systems endowed with
position-dependent effective mass have attracted, in recent years most
attention on behalf of physicists [1-6]. Effective mass Schr\"{o}dinger
equation was introduced by BenDaniel and Duke [1] in order to explain the
behavior of electrons in semiconductors. It have also many applications in
the fields of material sciences and condensed matter physics [7-9].

Exact solvability of the Schr\"{o}dinger equation was already discussed by
employing various techniques [10-20]. The underlying thoughts behind these
techniques might have different origins, while the applied technical
approaches are rather similar. The group-theoretical approaches are one of
these techniques and are useful for describing the bound-state problems
involving dynamical groups [13-20]. Originally, these approaches were used
to derive the Natanzon-class potentials [21] and their subclass as, an
example treated here, the Ginocchio potentials [22-24], using the new
concept of the potential group [17,18] which connects all states that have
the same energy but belong to different potential strengths.

In this paper, the \textit{conformal mappings} [25-27] are presented as a
means of generating the general Natanzon hypergeometric potentials (GNHP)
under the $\mathfrak{su}(1,1)$ group representation. The central observation
of the use of the conformal mappings consists in the fact that the variable $%
z$, as defined in the Natanzon-class potentials, varies in the interval $%
\left[ 0,1\right] $; this latter will be regarded as the radius of the unit
circle along the real-axis. This provides a systematic way for deriving
useful conformal mappings of the domain $\mathcal{D}$ lying in the $\xi -$%
plan onto the interior of the unit circle $\mathcal{D}_{\star }=\left[ 0,1%
\right] $ lying in the $z-$plan (see the Appendix).

The plan of the present paper is as follow. In section 2 we deduce the
expression of the effective potential using the differential realization of
the $\mathfrak{su}\left( 1,1\right) $ algebra, then exploiting the conformal
mappings lead to the general Natanzon hypergeometric potentials endowed with
the position-dependent mass. Section 3 deals with the generation of the
Ginocchio potentials in its hyperbolic and polynomial forms. The final
section will be devoted to discussions and an appendix was added, where the
mathematical details about conformal mappings will be presented.

\section{The $\mathfrak{su}(1,1)$\ algebra and Natanzon-class Potentials}

The $\mathfrak{su}\left( 1,1\right) $ Lie algebra consists of the three
generators $\mathcal{J}_{\pm },$ $\mathcal{J}_{0}$ satisfying the
commutation relations [13-20]%
\begin{equation}
\left[ \mathcal{J}_{+},\mathcal{J}_{-}\right] =-2\mathcal{J}_{0}\quad ;\quad %
\left[ \mathcal{J}_{0},\mathcal{J}_{\pm }\right] =\pm \mathcal{J}_{\pm }, 
\tag{1}
\end{equation}%
where $\mathcal{J}_{\pm }=\mathcal{J}_{\mp }^{\dag }$. They are related to
the Casimir operator as%
\begin{equation}
\mathcal{C=J}_{0}^{2}\mp \mathcal{J}_{0}-\mathcal{J}_{\pm }\mathcal{J}_{\mp
}.  \tag{2}
\end{equation}

Note that the eigenstates of $\mathcal{C}$ and $\mathcal{J}_{0}$, with
eigenvalues $\left\langle \mathcal{C}\right\rangle \equiv c=j\left(
j+1\right) $ and $\left\langle \mathcal{J}_{0}\right\rangle =j_{0}$, serve
as basis for the irreducible representation of $\mathcal{SU}\left(
1,1\right) $, and can be labelled $\left\vert j,j_{0}\right\rangle $. The
allowed values of $j_{0}$ are related to $j$ by [13,14]%
\begin{equation}
j_{0}=n+\frac{1}{2}+\sqrt{c+\frac{1}{4}},  \tag{3}
\end{equation}%
where $n=0,1,2,\ldots $ According to [19,20] the generators $\mathcal{J}%
_{\pm },$ $\mathcal{J}_{0}$ can be expressed in terms of the first-derivative%
\begin{eqnarray}
\mathcal{J}_{\pm } &=&\func{e}^{\pm i\varphi }\left[ \pm h\left( x\right)
\partial _{x}\pm g\left( x\right) +f\left( x\right) \mathcal{J}_{0}+c\left(
x\right) \right] ,  \TCItag{4.a} \\
\mathcal{J}_{0} &=&-i\partial _{\varphi },  \TCItag{4.b}
\end{eqnarray}%
where we have used the abbreviation $\partial _{\Sigma }=\frac{d}{d\Sigma }$%
, with $\Sigma =x,\varphi $. The Eqs. (1) and (4) provide the restrictions
which determine the shape of the functions $h\left( x\right) ,$ $c\left(
x\right) $ and $f\left( x\right) $ through the differential equations%
\begin{eqnarray}
f^{2}\left( x\right) -h\left( x\right) \partial _{x}f\left( x\right) &=&0, 
\TCItag{5.a} \\
h\left( x\right) \partial _{x}c\left( x\right) -f\left( x\right) c\left(
x\right) &=&0.  \TCItag{5.b}
\end{eqnarray}

By applying a variable transformation $h\left( x\right) \rightarrow h\left(
x\right) \frac{d\xi \left( x\right) }{dx}$ on (5), we obtain%
\begin{equation}
f\left( x\right) =\frac{1+a\xi ^{2}\left( x\right) }{1-a\xi ^{2}\left(
x\right) },\qquad c\left( x\right) =\frac{\delta \xi \left( x\right) }{%
1-a\xi ^{2}\left( x\right) },  \tag{6}
\end{equation}%
where $a$ and $\delta $ are constants of integration.

Inserting (6) into (2) and taking into account (4), we get%
\begin{equation}
\mathcal{C}=\frac{\xi ^{2}}{\xi ^{\prime 2}}\partial _{x}^{2}+\frac{\xi }{%
\xi ^{\prime }}\left[ 2g-\frac{\xi \xi ^{\prime \prime }}{\xi ^{\prime 2}}-%
\frac{2\xi ^{2}}{1-\xi ^{2}}\right] \partial _{x}+\frac{\xi }{\xi ^{\prime }}%
g^{\prime }+g^{2}-\frac{1+\xi ^{2}}{1-\xi ^{2}}g-\xi \frac{\left( \delta
+2j_{0}\xi \right) \left( 2j_{0}+\delta \xi \right) }{\left( 1-\xi
^{2}\right) ^{2}},  \tag{7}
\end{equation}%
where the prime denotes the derivative with respect to $x$. Eq.(7)
corresponds to the appropriate-parameter choice $a=1$.

On the other hand, the general form of the Hamiltonians introduced by von
Roos [2] for the spatially varying mass $M\left( x\right) =m_{0}m\left(
x\right) $, where $m\left( x\right) $ is a dimensionless mass, read%
\begin{equation}
\mathcal{H}_{VR}=\frac{1}{4}\left[ m^{\eta }\left( x\right) \widehat{\mathbf{%
p}}\ m^{\epsilon }\left( x\right) \widehat{\mathbf{p}}\ m^{\rho }\left(
x\right) +m^{\rho }\left( x\right) \widehat{\mathbf{p}}\ m^{\epsilon }\left(
x\right) \widehat{\mathbf{p}}\ m^{\eta }\left( x\right) \right] +V\left(
x\right) ,  \tag{8}
\end{equation}%
where $m_{0}=1$ and the restriction on the parameters $\eta ,$ $\epsilon $
and $\rho $ checks the condition $\eta +\epsilon +\rho =-1$. Here $\widehat{%
\mathbf{p}}\left( \equiv -i\hbar \partial _{x}\right) $ is the momentum. In
the natural units $\left( \hbar =c=1\right) $, the Hamiltonian $\mathcal{H}%
_{VR}$ becomes%
\begin{equation}
\mathcal{H}_{VR}=-\frac{1}{2m}\partial _{x}^{2}+\frac{m^{\prime }}{2m^{2}}%
\partial _{x}+\left( 1+\epsilon \right) \frac{m^{\prime \prime }}{4m^{2}}-%
\left[ \eta \left( \eta +\epsilon +1\right) +\epsilon +1\right] \frac{%
m^{\prime 2}}{2m^{3}}+V\left( x\right) .  \tag{9}
\end{equation}

By introducing the eigenfunctions [16]%
\begin{equation}
\psi _{\sigma }\left( x\right) =2\sigma m\left( x\right) \frac{\xi
^{2}\left( x\right) }{\xi ^{\prime 2}\left( x\right) }\phi \left( x\right) ,
\tag{10}
\end{equation}%
where $\sigma \in 
\mathbb{R}
$, the Hamiltonian (9) becomes%
\begin{eqnarray}
\mathcal{H}_{VR} &=&-\sigma \frac{\xi ^{2}}{\xi ^{\prime 2}}\partial
_{x}^{2}-\frac{\sigma \xi }{\xi ^{\prime }}\left[ 4+\frac{m^{\prime }\xi }{%
m\xi ^{\prime }}-\frac{4\xi \xi ^{\prime \prime }}{\xi ^{\prime 2}}\right]
\partial _{x}+\frac{2\sigma \xi }{\xi ^{\prime 2}}\left[ 3\xi ^{\prime
\prime }+\frac{\xi \xi ^{\prime \prime \prime }}{\xi ^{\prime }}-\frac{3\xi
\xi ^{\prime \prime 2}}{\xi ^{\prime 2}}\right]  \notag \\
&&+\frac{\sigma m^{\prime }\xi ^{2}}{m\xi ^{\prime 2}}\left[ \frac{2\left(
\xi \xi ^{\prime \prime }-\xi ^{\prime 2}\right) }{\xi \xi ^{\prime }}+\frac{%
\sigma \left( \epsilon -1\right) m^{\prime \prime }}{2m^{\prime }}-\left(
1+\eta \right) \left( \eta +\epsilon \right) \frac{m^{\prime }}{m}\right]
-2\sigma  \notag \\
&&+\frac{2\sigma m\xi ^{2}}{\xi ^{\prime 2}}V\left( x\right) .  \TCItag{11}
\end{eqnarray}

The Schr\"{o}dinger equation can be solved once equating it to the
eigenvalues equation of the Casimir invariant operator of the $\mathfrak{su}%
(1,1)$ algebra following [15]%
\begin{equation}
\left( \mathcal{H}_{VR}-E\right) \psi \left( x\right) =Z\left( x\right)
\left( \mathcal{C}-c\right) \psi \left( x\right) =0,  \tag{12}
\end{equation}%
where $Z\left( x\right) $ is some function to be determined. This
requirement provides the identities%
\begin{eqnarray}
Z\left( x\right) &=&-\sigma ,  \TCItag{13.a} \\
g\left( x\right) &=&\frac{2-\xi ^{2}\left( x\right) }{1-\xi ^{2}\left(
x\right) }-\frac{3\xi \left( x\right) \xi ^{\prime \prime }\left( x\right) }{%
2\xi ^{\prime 2}\left( x\right) }+\frac{m^{\prime }\left( x\right) \xi
\left( x\right) }{2m\left( x\right) \xi ^{\prime }\left( x\right) }. 
\TCItag{13.b}
\end{eqnarray}

Inserting $g\left( x\right) $, $g^{\prime }\left( x\right) $ and $%
g^{2}\left( x\right) $ as defined in (13.b) into (12), taking into
consideration (7) and (11), we end up with

\begin{equation}
V_{\text{eff}}\left( x\right) -E=\frac{2\delta j_{0}+\xi \left( \delta
^{2}+4j_{0}^{2}-1+2\delta j_{0}\xi \right) }{2m\xi \left( 1-\xi ^{2}\right)
^{2}}\xi ^{\prime 2}+\frac{c}{2m}\frac{\xi ^{\prime 2}}{\xi ^{2}}+\frac{3}{8m%
}\frac{\xi ^{\prime \prime 2}}{\xi ^{\prime 2}}-\frac{1}{4m}\frac{\xi
^{\prime \prime \prime }}{\xi ^{\prime }}+\mathcal{V}_{\text{m}}^{\left(
\eta ,\epsilon \right) }\left( x\right) ,  \tag{14}
\end{equation}%
where%
\begin{equation}
\mathcal{V}_{\text{m}}^{\left( \eta ,\epsilon \right) }\left( x\right) =%
\frac{m^{\prime 2}}{8m^{3}}\left[ \left( 1+2\eta \right) ^{2}+4\epsilon
\left( 1+\eta \right) \right] -\frac{\epsilon m^{\prime \prime }}{4m^{2}}, 
\tag{15}
\end{equation}

Now to derive the GNHP and the different steps of calculations that can
arise, we introduce first a transformation deduced from the conformal
mappings and discussed in the appendix%
\begin{equation}
\xi \left( x\right) =\frac{1+i\sqrt{z\left( x\right) }}{1-i\sqrt{z\left(
x\right) }},  \tag{16}
\end{equation}%
followed by replacing $z\left( x\right) \rightarrow -z\left( x\right) $,
where the variable function $z\left( x\right) $ varies in the interval $%
\left[ 0,1\right] $. Then (14) becomes 
\begin{equation}
E-V_{\text{eff}}\left( x\right) =\frac{pz-q-1}{4z^{2}\left( 1-z\right) }%
\frac{z^{\prime 2}}{2m}-\frac{c}{z\left( 1-z\right) ^{2}}\frac{z^{\prime 2}}{%
2m}-\frac{3}{8m}\frac{z^{\prime \prime 2}}{z^{\prime 2}}+\frac{1}{4m}\frac{%
z^{\prime \prime \prime }}{z^{\prime }}-\mathcal{V}_{\text{m}}^{\left( \eta
,\epsilon \right) }\left( x\right) ,  \tag{17}
\end{equation}%
where%
\begin{eqnarray}
p &\equiv &t-1=\left( \frac{\delta -2j_{0}}{2}\right) ^{2}-1,  \TCItag{18.a}
\\
q+1 &\equiv &r-1=\left( \frac{\delta +2j_{0}}{2}\right) ^{2}-1, 
\TCItag{18.b}
\end{eqnarray}

Without loss of generality, let us assume that the function $z\left(
x\right) $ is related to a certain \textit{generating function}, namely, $%
\mathfrak{S}\left( x\right) $ by%
\begin{equation}
\mathfrak{S}\left( x\right) =\frac{z^{\prime 2}\left( x\right) }{2m\left(
x\right) },  \tag{19}
\end{equation}

By performing a formal derivative of (17) taking into account (19), we obtain%
\begin{equation}
\mathfrak{S}\left( x\right) =-\frac{4z^{2}\left( x\right) }{\frac{4z\left(
x\right) }{\left( 1-z\left( x\right) \right) ^{2}}\partial _{E}c-\frac{%
z\left( x\right) }{1-z\left( x\right) }\partial _{E}p+\frac{1}{1-z\left(
x\right) }\partial _{E}q}.  \tag{20}
\end{equation}

Henceforth, we assume that the derivatives of the coefficients $p$, $q$ and $%
c$ with respect of $E$ in (20) are constant, which requires that the
coefficients are linear with respect to $E$ [13-15]. In terms of these
settings, the coefficients become%
\begin{eqnarray}
c\left( E\right) &=&-c_{0}E+a_{c},  \TCItag{21.a} \\
p\left( E\right) &=&-p_{0}E+a_{p},  \TCItag{21.b} \\
q\left( E\right) &=&-q_{0}E+a_{q},  \TCItag{21.c}
\end{eqnarray}%
where $c_{0}$, $p_{0}$, $q_{0}$, $a_{c}$, $a_{p}$ and $a_{q}$ are six real
parameters. A straightforward algebraic manipulation permits to recast the
generating function $\mathfrak{S}\left( x\right) $ through the differential
equation in $z\left( x\right) $%
\begin{equation}
\mathfrak{S}\left( x\right) \equiv \frac{z^{\prime 2}\left( x\right) }{%
2m\left( x\right) }=\frac{4z^{2}\left( x\right) \left[ 1-z\left( x\right) %
\right] ^{2}}{\mathcal{R}\left[ z\left( x\right) \right] },  \tag{22}
\end{equation}%
where%
\begin{equation}
\mathcal{R}\left[ z\left( x\right) \right] =p_{0}z^{2}\left( x\right)
+\left( 4c_{0}-p_{0}-q_{0}\right) z\left( x\right) +q_{0}.  \tag{23}
\end{equation}

Substituting now Eqs. (19) and (21) into (17) we obtain%
\begin{equation}
V_{\text{eff}}\left( x\right) =\frac{a_{p}z^{2}-\left(
a_{p}+a_{q}-4a_{c}+1\right) z+a_{q}+1}{\mathcal{R}\left[ z\left( x\right) %
\right] }+\frac{5}{32m}\left( \frac{\mathfrak{S}^{\prime }}{\mathfrak{S}}%
\right) ^{2}-\frac{1}{8m}\frac{\mathfrak{S}^{\prime \prime }}{\mathfrak{S}%
^{\prime }}+\frac{m^{\prime }}{16m^{2}}\frac{\mathfrak{S}^{\prime }}{%
\mathfrak{S}}+\mathcal{U}_{\text{m}}^{\left( \eta ,\epsilon \right) }\left(
x\right) ,  \tag{24}
\end{equation}%
where%
\begin{equation}
\mathcal{U}_{\text{m}}^{\left( \eta ,\epsilon \right) }\left( x\right) =%
\left[ \frac{4\left( 1+2\eta \right) ^{2}+16\epsilon \left( 1+\eta \right) +5%
}{32}\right] \frac{m^{\prime 2}}{m^{3}}-\frac{2\epsilon +1}{8}\frac{%
m^{\prime \prime }}{m^{2}}.  \tag{25}
\end{equation}

Knowing (22), $\mathfrak{S}^{\prime }\left( x\right) $ and $\mathfrak{S}%
^{\prime \prime }\left( x\right) $ can be expressed in terms of $z\left(
x\right) $ leading, after long and straightforward algebras, to express the
effective potential (24) in the form%
\begin{eqnarray}
V\left( x\right) &=&\frac{a_{p}z^{2}-\left( a_{p}+a_{q}-4a_{c}+1\right)
z+a_{q}+2}{\mathcal{R}\left[ z\left( x\right) \right] }  \notag \\
&&+\left[ p_{0}+\frac{\left( 4c_{0}-q_{0}\right) \left( 2z-1\right) +p_{0}}{%
z\left( z-1\right) }-\frac{5\Delta }{4\mathcal{R}\left[ z\left( x\right) %
\right] }\right] \left[ \frac{z\left( z-1\right) }{\mathcal{R}\left[ z\left(
x\right) \right] }\right] ^{2},  \TCItag{26}
\end{eqnarray}%
where $V\left( x\right) =V_{\text{eff}}\left( x\right) -\mathcal{U}_{\text{m}%
}^{\left( \eta ,\epsilon \right) }\left( x\right) $ and $\Delta =\left(
4c_{0}-p_{0}-q_{0}\right) ^{2}-4p_{0}q_{0}$.

We recognize in (26) the expression of the \textit{general} \textit{Natanzon
hypergeometric potentials }[10,15,21]. The bound-states spectra can be
determined from (18), taking into account (3), following the identity%
\begin{equation}
\sqrt{q+2}-\sqrt{p+1}-\sqrt{4c+1}\equiv 2n+1.  \tag{27}
\end{equation}

\section{A particular example : Ginocchio potentials}

Probably the most-known member of the Natanzon-class is the Ginocchio
potentials [22,23] which has as an important feature the possibility to be
reduced to the P\"{o}schl-Teller potential [24] in the one-dimension case
and to the Generalized P\"{o}schl-Teller potential in the radial case. It is
a perfect example of "\textit{implicit}"\ potentials; i.e. it is expressed
in terms of a function $z\left( x\right) $ which is known only in the
implicit form $x\left( z\right) $. Consequently, the bound-states spectra
are then given by a more complicated form. Setting the appropriate-parameter
choices $c_{0}=\frac{1}{4\gamma ^{4}},$ $a_{c}=-\frac{1}{4},$ $p_{0}=\frac{%
1-\gamma ^{2}}{\gamma ^{4}},$ $a_{p}=\left( j+\frac{1}{2}\right) ^{2}-1,$ $%
q_{0}=0,$ and $a_{q}=-\frac{7}{4}$, and combining (22) to (19), we obtain
the dimensionless mass integral%
\begin{equation}
\mu \left( x\right) \equiv \dint\limits^{x}dy\sqrt{2m\left( y\right) }=\frac{%
1}{2\gamma ^{2}}\dint\limits^{z\left( x\right) }\frac{ds\left( x\right) }{%
1-s\left( x\right) }\sqrt{1-\gamma ^{2}+\frac{\gamma ^{2}}{s\left( x\right) }%
}.  \tag{28}
\end{equation}

By defining a new variable transformation $s\left( x\right) =\tanh
^{2}u\left( x\right) $, (28) is reduced to%
\begin{equation}
\mu \left( x\right) =\frac{1}{\gamma ^{2}}\dint\limits^{z\left( x\right)
}du\left( x\right) \frac{\sqrt{\gamma ^{2}+\sinh ^{2}u\left( x\right) }}{%
\cosh u\left( x\right) },  \tag{29}
\end{equation}%
where it is impossible to get $z\left[ \mu \right] \left( x\right) $, the
solution of (29) in closed form; rather only an implicit $\mu \left[ z\right]
\left( x\right) $ function can be determined given by%
\begin{equation}
\mu \left( x\right) =\frac{1}{\gamma ^{2}}\func{arctanh}\left[ \frac{\sinh
z\left( x\right) }{\sqrt{\gamma ^{2}+\sinh ^{2}z\left( x\right) }}\right] +%
\frac{\sqrt{\gamma ^{2}-1}}{\gamma ^{2}}\arctan \left[ \frac{\sqrt{\gamma
^{2}-1}\sinh z\left( x\right) }{\sqrt{\gamma ^{2}+\sinh ^{2}z\left( x\right) 
}}\right] .  \tag{30}
\end{equation}

By inserting the parameters mentioned above in (26), taking into
consideration (23), we end up obtaining the expression of the Ginocchio
potentials either in its "\textit{hyperbolic form}"\ [23], given by%
\begin{equation}
V_{\text{hyp.}}\left( x\right) =-\gamma ^{4}\frac{j\left( j+1\right) -\gamma
^{2}+1}{\gamma ^{2}+\sinh ^{2}z\left( x\right) }-\frac{3}{4}\frac{\gamma
^{4}\left( 3\gamma ^{2}-1\right) \left( \gamma ^{2}-1\right) }{\left[ \gamma
^{2}+\sinh ^{2}z\left( x\right) \right] ^{2}}+\frac{5}{4}\frac{\gamma
^{6}\left( \gamma ^{2}-1\right) ^{2}}{\left[ \gamma ^{2}+\sinh ^{2}z\left(
x\right) \right] ^{3}},  \tag{31}
\end{equation}%
or in its "\textit{polynomial form}"\ [18,22]%
\begin{equation}
V_{\text{poly.}}\left( x\right) =\left[ -\gamma ^{2}j\left( j+1\right) +%
\frac{1-\gamma ^{2}}{4}\left\{ 2-\left( 7-\gamma ^{2}\right) y^{2}\left(
x\right) +5\left( 1-\gamma ^{2}\right) y^{4}\left( x\right) \right\} \right]
\left( 1-y^{2}\left( x\right) \right) ,  \tag{32}
\end{equation}%
once the variable transformation%
\begin{equation}
y\left( x\right) =\frac{\sinh z\left( x\right) }{\sqrt{\gamma ^{2}+\sinh
^{2}z\left( x\right) }},  \tag{33}
\end{equation}%
is introduced in (31), where $-1\leq y\leq 1$.

Proceeding now to squaring (27), then it is easy to obtain the expression of
the bound-states spectra [18,22,23]%
\begin{equation}
E_{n}=-\left[ \sqrt{\left( 1-\gamma ^{2}\right) \left( 2n+\frac{1}{2}\right)
^{2}+\gamma ^{2}\left( j+\frac{1}{2}\right) ^{2}}-\left( 2n+\frac{1}{2}%
\right) \right] ^{2},  \tag{34}
\end{equation}%
where $n=0,1,2,\ldots ,\left[ j\right] $.$\footnote[1]{$\left[ j\right] $
means the integer part of $j$.}$

\section{Conclusion}

The conformal mappings have been used to generate the general Natanzon
hypergeometric potentials (GNHP) endowed with a position-depend mass in the
framework of the $\mathfrak{su}\left( 1,1\right) $ group representation, and
as an example, we have derived the Ginocchio potentials and corresponding
bound-state spectra as well. Here the particular interest carried upon the
conformal mappings is due essentially to the fact that the variable $z$, as
defined in Natanzon-class potentials, belongs to the interval $\left[ 0,1%
\right] $, this led us to establish the connection between $\xi \in \mathcal{%
D}$ and $z\in \mathcal{D}_{\star }$. To be more precise, it has been shown
that the GNHP can be deduced under the linear-fractional function (16)
specified by a conformal mapping; this means that the function $z$, in (16),
specifies a mapping under which the points of the real axis $\func{Im}\Omega
=\func{const}$, where $\xi =\exp \left[ 2i\Omega \right] \footnote[1]{%
For discussion of such function, cf. the formulas (A.1) and (A.2) in the
appendix.}$, are one-sheeted correspondence with the points of the contour $%
\left\vert z\right\vert \leq 1$. Consequently, this function performs a
conformal mapping of the upper half-plane onto the interior of the unit
circle.

The conformal mappings can be considered as one of powerful methods of
generating the exactly (may be also quasi-exactly) potentials from a
different perspective using only the geometric aspects, aspects that will be
useful in visualizing the connection between the domains.

\section{Appendix : Discussion of conformal mappings}

As outlined in the introduction, our immediate purpose in this appendix is
to see how the transformation given in (16) can be deduced from a \textit{%
conformal mappings} specified by some elementary and analytic functions.

Given an analytic function $w=f\left( z\right) $ in a domain $\mathcal{D}$,
to each point $z\in \mathcal{D}$ there corresponds a definite point on the
complex-plan of the variable $w\in \mathcal{D}_{\star }$. If this
correspondence between $z$ and $w$ is one-to-one, the function $w=f\left(
z\right) $ is called \textit{one-sheeted}. In case of such correspondence,
we say that there is a \textit{mapping of the domain }$\mathcal{D}$\textit{\
onto the domain }$\mathcal{D}_{\star }$. The point $w\in \mathcal{D}_{\star
} $ is called the image of the point $z\in \mathcal{D}$ and the point $z$ is
called the original of the point $w$ [25-27].

Let the domain $\mathcal{D}$ belongs to the $z-$plan. The introduction of
the transformation%
\begin{equation}
\mathfrak{z}=2iz,  \tag{A1}
\end{equation}%
where $\mathfrak{z}=x_{1}+iy_{1}$, allows to perform a pure rotation through
an angle $\frac{\pi }{2}$ and a double dilatation of our domain. Then we
make up the exponential function%
\begin{equation}
\mathcal{Z}=\exp \left[ \mathfrak{z}\right] .  \tag{A2}
\end{equation}

From $\mathcal{D}$, let us choose $\func{Re}z$ defined in the band : $-\frac{%
\pi }{4}\leq \func{Re}z\leq \frac{\pi }{4}$, with the correspondence of
three neighborhood points $f\left( \pm \frac{\pi }{4}\right) =\pm 1$, $%
f\left( i\infty \right) =i$ (here $i\infty $ indicates the point located at
infinity in the direction of the imaginary axis of the $z-$plan).
Consequently, the transformations (A1) and (A2) transform the band $-\frac{%
\pi }{2}\leq \func{Im}\mathfrak{z}\leq \frac{\pi }{2}$, on which the
function (A1) transforms $\func{Re}z$, onto the upper half-plan $\func{Re}%
\mathcal{Z}>0$ (In fact as $\mathcal{Z}=\func{e}^{x_{1}}\func{e}^{iy_{1}}$,
then $\left\vert \mathcal{Z}\right\vert =\func{e}^{x_{1}}$ varies from $0$
to $\infty $ and $\arg \mathcal{Z}=y_{1}$ from $-\frac{\pi }{2}$ to $\frac{%
\pi }{2}$). It remains to transform this upper half-plan onto the unit
circle so that the points $\mathcal{Z}=\pm i,0$ which correspond
respectively to the points $z=\pm \frac{\pi }{4},i\infty $ have as images
the points $w=\pm 1,i$. This problem can be solved by giving the
transformation [25]%
\begin{equation}
\frac{w-w_{1}}{w-w_{3}}\frac{w_{2}-w_{3}}{w_{2}-w_{1}}=\frac{\mathcal{Z}%
-z_{1}}{\mathcal{Z}-z_{3}}\frac{z_{2}-z_{3}}{z_{2}-z_{1}},  \tag{A3}
\end{equation}%
leading to the \textit{linear-fractional} (\textit{homographic}) functions%
\begin{equation}
w=\frac{1}{i}\frac{\mathcal{Z-}1}{\mathcal{Z+}1}.  \tag{A4}
\end{equation}

Substituting in (A4) the expressions of (A1) and (A2) we find the solution
of the problem as%
\begin{equation}
w=\frac{1}{i}\frac{\func{e}^{2iz}\mathcal{-}1}{\func{e}^{2iz}\mathcal{+}1}%
\equiv \tan z,  \tag{A5}
\end{equation}%
where $-1\leq \func{Re}w\leq 1$. Finally, taking into account the inverse of
(A4), we have%
\begin{equation}
\mathcal{Z}=\frac{1+iw}{1-iw},  \tag{A6}
\end{equation}%
and after performing the transformation $w\rightarrow v=\sqrt{w}$ (i.e. $%
\rho =r^{1/2}$ and $2\varphi =\theta $) one ends up obtaining the desired
expression (16). We now have two points in the $v-$plan corresponding to one
point in the $w-$plan. The important point here is that we can make the
function $v$ a single-valued function instead of a double-valued function if
we agree to restrict $\theta $ to a range such as $0\leq \theta \leq 2\pi $
[27]. This may be done by agreeing never to cross the line $\theta =0$ in
the $w-$plan. Such a line of demarcation is known as a \textit{cut line}, of
which the mean purpose consists to restrict the argument of $w$, leading to
write $\func{Re}w\in \left( 0,1\right] $. Then $w=0$ is a \textit{branch
point}, which brings to conclude that $v=\sqrt{w}$ not being analytic at the
point $w=0$.

\end{document}